\documentclass[sigconf]{acmart}

\usepackage{subcaption}
\usepackage{multirow}
\usepackage{graphicx}
\usepackage{caption}
\usepackage{booktabs}
\usepackage{epstopdf}

\copyrightyear{2022}
\acmYear{2022}
\setcopyright{rightsretained}
\acmConference[CHI '22 Extended Abstracts]{CHI Conference on Human
Factors in Computing Systems Extended Abstracts}{April 29-May 5,
2022}{New Orleans, LA, USA}
\acmBooktitle{CHI Conference on Human Factors in Computing Systems
Extended Abstracts (CHI '22 Extended Abstracts), April 29-May 5, 2022,
New Orleans, LA, USA}
\acmDOI{10.1145/3491101.3519658}
\acmISBN{978-1-4503-9156-6/22/04}

\begin{document}

\title{Correlation between Unconscious Mouse Actions and Human Cognitive Workload}

\author{Go-Eum Cha}
\email{cha20@purdue.edu}
\orcid{0000-0002-9812-636X}
\affiliation{%
  \institution{Purdue University}
  \streetaddress{610 Purdue Mall}
  \city{West Lafayette}
  \state{Indiana}
  \country{USA}
  \postcode{47907}
}
\author{Byung-Cheol Min}
\authornotemark[1]
\email{minb@purdue.edu}
\orcid{0000-0001-6458-4365}
\affiliation{%
  \institution{Purdue University}
  \streetaddress{610 Purdue Mall}
  \city{West Lafayette}
  \state{Indiana}
  \country{USA}
  \postcode{47907}
}

\titlenote{\Large Accepted ACM CHI 22' Late-Breaking Work (LBW). This manuscript is a pre-print version.}
\renewcommand{\shortauthors}{Go-Eum Cha and Byung-Cheol Min}

\begin{abstract}
  Unconscious behaviors are one of the indicators of the human perception process from a psychological perspective. As a result of perception responses, hand gestures show behavioral responses from given stimuli. Mouse usages in Human-Computer Interaction (HCI) show hand gestures that individuals perceive information processing. This paper presents an investigation of the correlation between unconscious mouse actions and human cognitive workload. We extracted mouse behaviors from a Robot Operating System (ROS) file-based dataset that user responses are reproducible. We analyzed redundant mouse movements to complete a dual $n$-back game by solely pressing the left and right buttons. Starting from a hypothesis that unconscious mouse behaviors predict different levels of cognitive loads, we statistically analyzed mouse movements. We also validated mouse behaviors with other modalities in the dataset, including self-questionnaire and eye blinking results. As a result, we found that mouse behaviors that occur unconsciously and human cognitive workload correlate.
\end{abstract}



\maketitle

\section{Introduction}
Human operators play an essential part in Human-Computer Interaction (HCI) to achieve better performance. Operators produce activities with standard input devices, a mouse and a keyboard, and retrieve outputs through their monitors to complete tasks that can be solved by interactions. The usage through input devices is the target of flourishing the quality of interaction between two different subjects, mainly understanding human states is the top priority. Human cognitive and affective states influence the performance of operators \cite{carroll1997human}. If operators are allocated several tasks persistently, the consecutive tasks may provoke leakage of cognitive capabilities \cite{debie2019multimodal}, which would cause adverse effects on task completion. Researchers have carried out studies to measure cognitive workload in behavioral, physiological, and neurophysiological methods.  

Memory in the cognitive process arguably explains the ability that previously acquired information is retrievable over time. \cite{cowan2008differences} described working memory as a kind of short-term memory. However, attainable human memory has a capacity limit, which leads people to choose tactical ways to maintain the information. Numerous human reactions have been investigated to measure a person's workload from demonstrated research results that internal states of human and their reactions correlates \cite{morsella2011unconscious, rheem2018use}. Grounded from top-down approach that indicates humans concentrate on task-relevant stimuli selectively, and irrelevant ones to tasks will be suppressed otherwise \cite{gazzaley2012top}, explicit variations of expressive behaviors, such as motor behaviors, have been elicited to be matched with cognitive load. 

Human hand gestures have been used as a measure to gauge a person's perception response from stimuli imitating genuine human-computer interaction. Mouse dynamics on task completion have mainly been scrutinized \cite{grimes2015mind, rheem2018use, freihaut2021using, witte2021measuring}, and mutually meet agreements that mouse usage during task completion indicates the intensity of cognitive burden on humans. Higher cognitive load humans intrinsically undergo, less active mouse usage, such as pixel variation, speed, and click pressure occurs \cite{rheem2018use, grimes2015mind, witte2021measuring}. However, previous studies required participants to drag a mouse actively when completing given laboratory tasks. The task-irrelevant motor behaviors \cite{lau2007unconscious} have not been deeply investigated. The task-related motor behaviors can be spotted during completion, but it is ambiguous whether the hand behaviors were derived directly from tasks or the internal process.


Studies on unintended behavioral reactions and cognitive loads are scarce. When human's internal states change, facial expressions or motor behaviors occur because of given stimuli \cite{morsella2011unconscious, zimmermann2003affective}. The changes happen unconsciously, but where the events can influence cognition or motor outcomes. These behaviors voluntarily happen as self-motor behaviors. 
For example, touching a hot surface makes us hands off right away. The process of human perception can also be observed as motor behavior beyond conscious awareness. Visual stimulation derives unconscious actions during humans process complex visual information \cite{maruya2007voluntary}. Finger tapping when listening to music appears as a process of auditory stimulation \cite{maes2014action}. Given this finding, the question arises as to whether redundant hand gestures can represent a person's cognitive load.

In this paper, we study the unconscious mouse movements of multiple operators performing the dual $n$-back test. We investigate physical movements and self-questionnaire data between mouse usage and cognitive load when humans do not need to move their hands. We also analyze the attained data from the game, presuming humans would make redundant mouse movements when cognitive load gradually increases. We expect that the task-irrelevant mouse behaviors enable us to predict different mental demands. 

\section{RELATED WORK}
\label{literature}
Mouse dynamics in HCI have been extensively explored as one of the human internal state indicators. Data from mouse dynamics vary, such as pixel variation, speed, and response time. \cite{rheem2018use} conducted a user study to assess cognitive load providing participants experiments clicking distant and different sized circles. The researchers found that the reaction time and movement velocity in the highest cognitive condition deviates from the other two lower levels. Similarly, \cite{grimes2015mind} made participants compare numbers and click the right button on the screen, and the level increased memorizing a few numbers. They found that longer distances and slower movements happened when higher cognitive loads were derived. Given the previous studies, we have selected mouse dynamics to scrutinize the relationship between human cognition and unconscious actions through a game that can derive cognitive load. 

The field of human workload estimation has been extensively studied over decades. Cognitive load is ambiguous that cannot quantitatively measure the exact amount, which led researchers to find equations to derive the load \cite{sampei2016mental} utilizing NASA Task Load Index (NASA-TLX) \cite{hart1988development}. The situation that humans take part in varies, including aircraft \cite{borghini2014measuring, peruzzini2019transdisciplinary}, automobile \cite{benedetto2011driver, fridman2018cognitive}, and human-robot interaction \cite{rabby2019effective}. In various situations, no dominant methods were presented in the previous studies. Researchers have adopted single or multi modalities to predict mental load in various ways. Physiological or neurophysiological sensors are getting popular, while some researchers have raised that a non-intrusive system does not bother human reactions to measure mental load \cite{peruzzini2019transdisciplinary, cech2016real}. Behavioral measures, such as keystrokes or mouse tracking \cite{evans2017multi, freihaut2021tracking} have been selected to predict mental load. The difference between behavioral measures and sensor-based ones is whether the measures perform given tasks directly. Sensor responses occur as physical reactions caused by the cognition process, while humans follow out a course of action connected to behavioral reaction. Load estimation varies from personal behavior patterns as well. In the perspective of behavioral measures, workload estimation is mainly related to reactions that are directly associated with information processing and task completion. 
The $n$-back task has been widely adopted to derive cognitive load evaluate working memory capabilities \cite{jaeggi2010relationship, lawlor2016dual, he2019high}. The dual $n$-back gives two kinds of stimuli, visually and auditorily, to match previous $n$-steps cue to the most recent one. 

Unconscious actions in a psychological perspective are behavioral responses resulting from cognition and motor process that do not participate in conscious perception \cite{morsella2011unconscious}. Behaviors that people do not even recognize about themselves can also represent the cognitive process. The unconscious or voluntary actions can be indicators of the stimulation process. For example, a human shows processing states from visual motivation to their motor behaviors beyond human awareness \cite{maruya2007voluntary}. Not only can visual stimulus derive human motor control, but auditory motives provoke physical activities. Small movements that appear when people listen to music are also caused by cognitive processes that occur intrinsically. Researchers analyze these actions in terms of social interactions as well as expressions \cite{maes2014action}, concluding body movements reflect the cognitive process that can be monitored beyond consciousness.

The background of cognitive workload estimation and unconscious action brought that task-irrelevant behaviors may predict the degree of mental workload. To measure the load using these unconscious actions and eye-related actions when completing tasks have been selected. We chose mouse movements, considering HCI situations, to see if unaware behaviors that can be monitored through body movements are also cues to estimate cognitive burden. As mouse activities have drawn attention to investigate affective states and stress in an unobtrusive way \cite{freihaut2021tracking}, it is uncertain that unconscious mouse behaviors have a similar effect as task-related behaviors do.

\subsection{Hypothesis}

When experiencing cognitive load, mouse behaviors that are not relevant to task completion, or unconscious activity of the dual $n$-back games will change following behaviors based on the three different levels: Frequency of movements (H1a), Moving duration (H1b), and Movement position changes; pixels (H1c).

\section{Dataset} \label{dataset}

\begin{figure}
\centering
\includegraphics[width=0.96\linewidth]{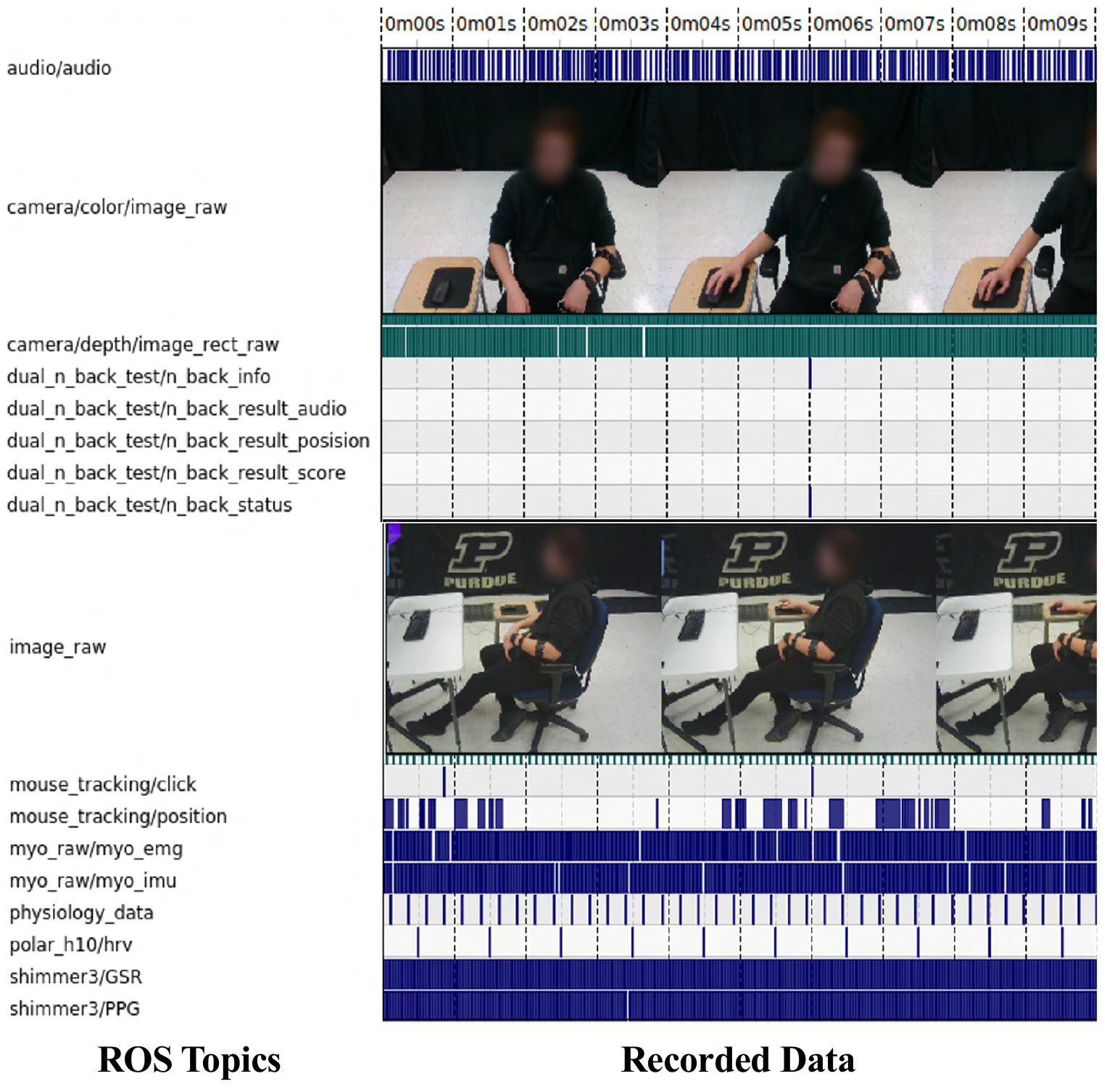} 
\caption{An example of the ROSBag files in the workload test in the affective dataset \cite{jo2020rosbag}. The blue bars represent recorded data with time frame.}
\label{img:dataset}
\Description{ROSBag data with videos, mouse behavior, physiological sensors have been presented. ROS topics are listed, and data segments are presented along with topics. The video topics have representative frames.}
\end{figure}

We utilized the workload test set in the affective dataset \cite{jo2020rosbag} that collected human behavioral, physiological, and self-rating responses. The dataset includes 30 participants (11 females and 19 males: the age ranges from 18 to 37; mean: 25.1; std: 4.497). As Fig.~\ref{img:dataset} depicts how the ROSBag dataset looks like, the dataset provided a human facial video, recorded mouse data, and data of physiological sensors during experiments while playing the dual $n$-back games. 
All streams of data can be retrieved by replaying pre-defined topics. For example, as shown in Fig.~\ref{img:dataset}, mouse position changes during the experiment can be retrieved by using the ROS topic named \textit{/mouse\_tracking/position}. The ROS topic consists of structured data of coordinates with their timestamps. The dataset also provides the NASA-TLX self-questionnaire obtained after the dual $n$-back game rounds. Consequently, the dataset allowed us to obtain the externally observable participant's behaviors and the self-questionnaire results of cognitive load. 

However, for the analysis of mouse behaviors, the data of 27 individuals were used because the data of three participants was only partially recorded across all topics. Data that can only be retrieved in part may compromise data integrity, hence three participants' data were excluded. 
The data of 16 participants among 27 participants also included facial and figural videos of participants who did not wear eyeglasses. Glares interfered with analysis on eye behaviors since the OpenPose \cite{openposegithub} could not successfully detect eye features in facial video streams. Therefore, the data of 16 participants were utilized in further validation steps. 

\begin{figure}
\centering
\includegraphics[width=0.95\linewidth]{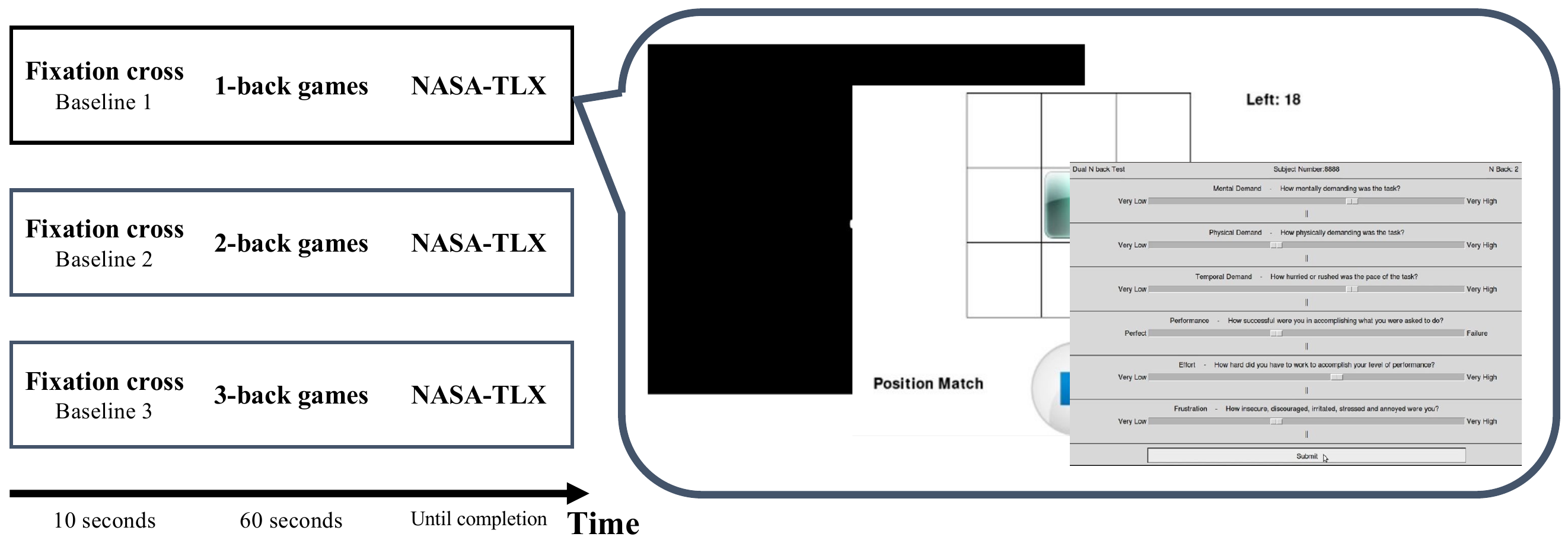}
\caption{The procedure of the user experiment with the dual $n$-back task used for this study. Watch the supplementary video for more detail.}
\label{img:experiment_procedure}
\Description{The procedure of the experiment is drawn in the table and screen captures. The group of 'Fixation cross,' 'n-back game,' and 'NASA-TLX' is repeated three times, and one rectangle figure contains screenshots of the experiment indicating each group went the same procedure.}
\end{figure}

The game exposes a cue during 3000 milliseconds, requiring players to remember the subject's location or numbers that the participants saw or heard $n$ turns back. The dual 1-back game is represented in a low level, dual 2-back as a medium, and dual 3-back as a high one. Fig.~\ref{img:experiment_procedure} depicts the entire process of the user experiment with the dual $n$-back task. Participants firstly saw a fixation cross ten seconds before starting actual workload inducement. After being exposed to the cross, the game continues until participants complete one round for 60 seconds. When the 60 seconds of the game completed, participants filled the NASA-TLX to answer mental demand, physical demand, temporal demand, performance, effort, and frustration levels. 

\begin{figure}
  \centering
  \begin{subfigure}[]{0.49\linewidth}
         \centering
         \includegraphics[width=\linewidth]{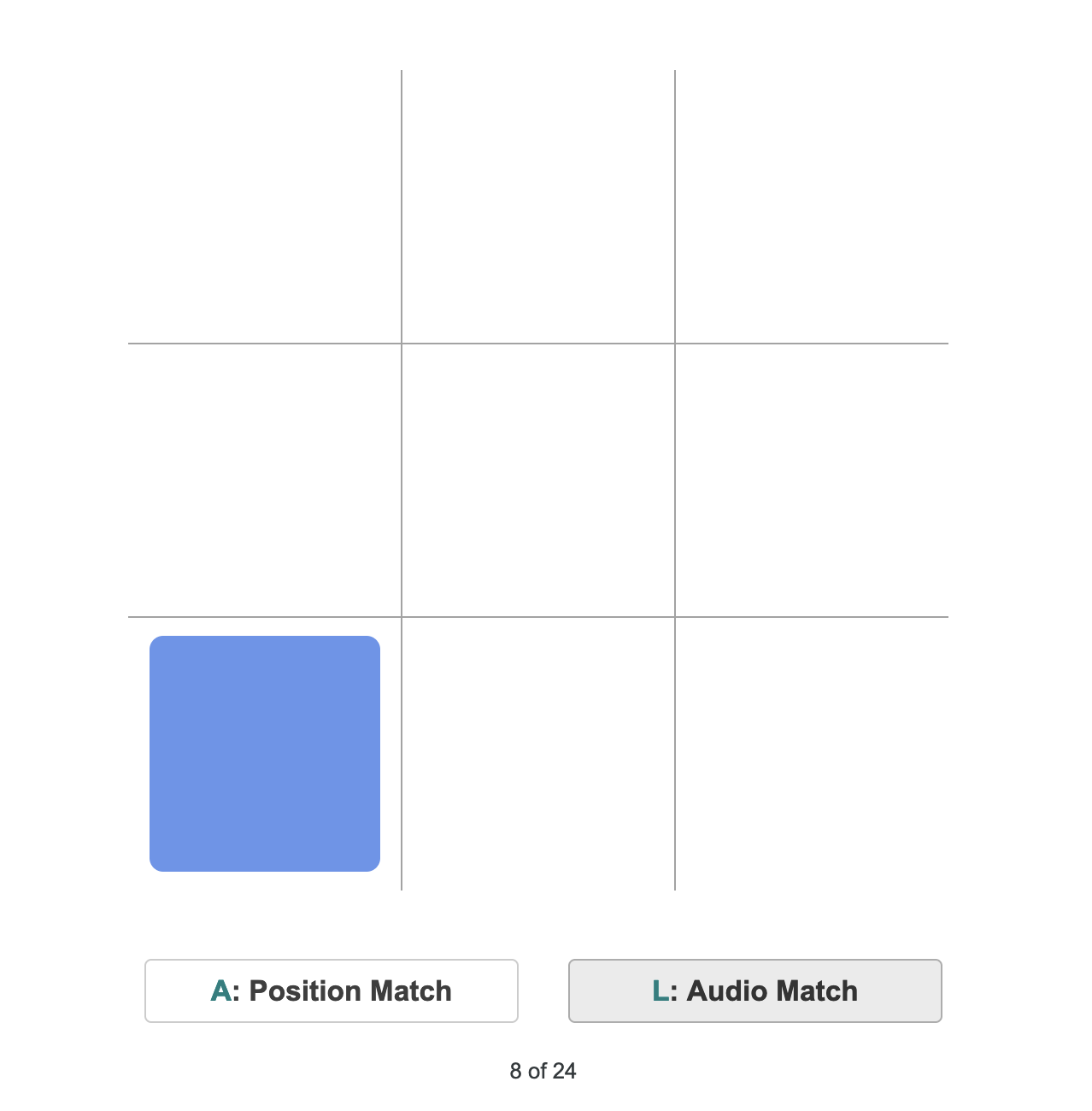}
         \caption{}
     \end{subfigure}
     \begin{subfigure}[]{0.50\linewidth}
     \vspace{6mm}
         \centering
         \includegraphics[width=\linewidth]{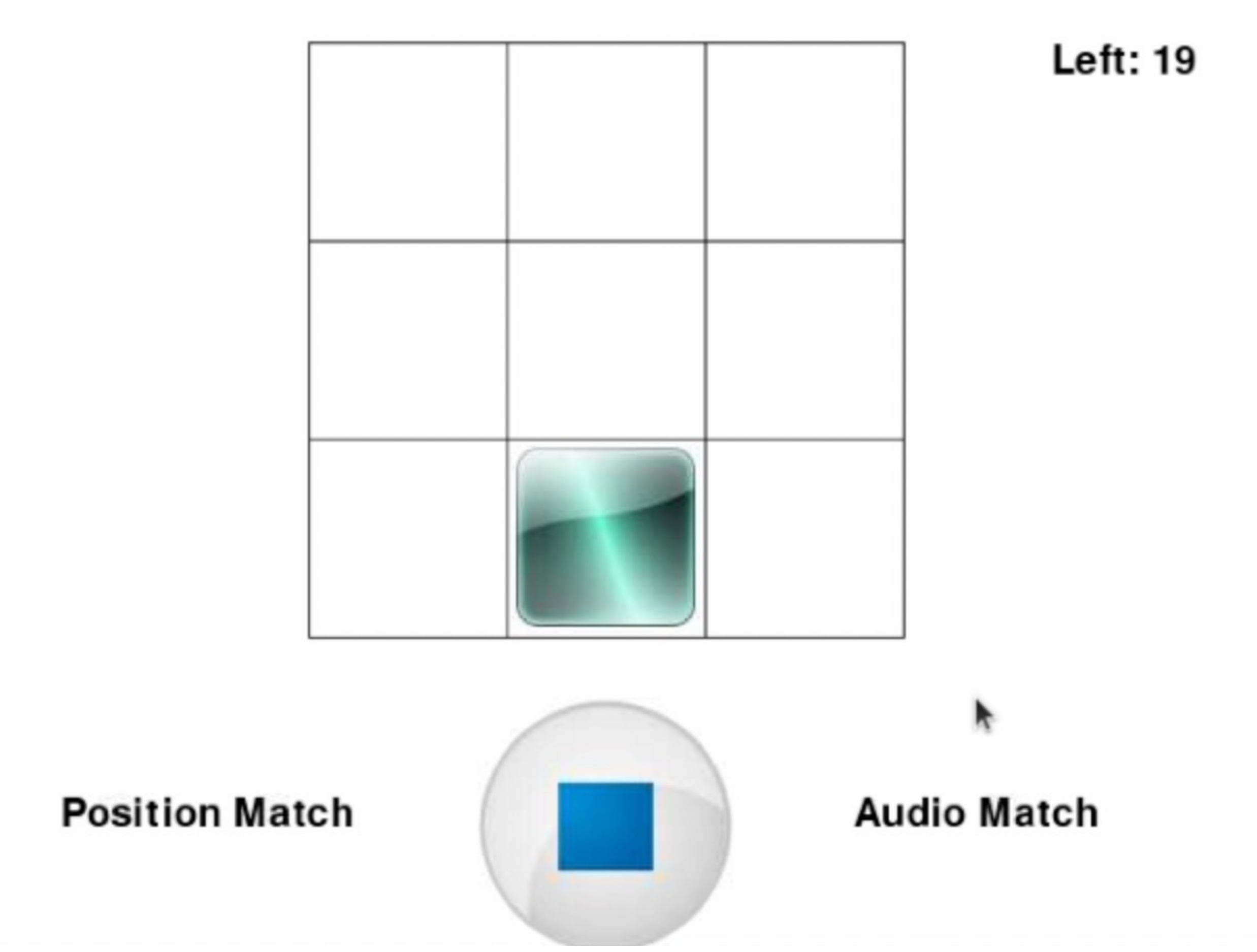}
         \label{img:n-back-gui}
         \caption{}
     \end{subfigure}
  \caption{Comparisons of the dual $n$-back games: (a) the traditional dual $n$-back game where the participant can answer two matches by clicking a button on GUI or hitting a keyboard (A or L) \cite{brainscale}, and (b) the dual $n$-back game \cite{jo2020rosbag} used for this study where the participant can only answer by pressing a mouse button.}
  \label{img:gui_comparison}
  \Description{Two different types of the dual $n$-back GUIs are presented. One GUI on the left has two click buttons to indicate audio or visual matches by moving the mouse cursor. The other GUI is our GUI that does not need to move the cursor while playing the dual n-back game.}
\end{figure}

Unlike the traditional way of answering two matches by clicking a button on the Graphic User Interface (GUI) or hitting keyboard \cite{jaeggi2010relationship, lawlor2016dual} as shown in Fig.~\ref{img:gui_comparison} (a), this study was designed to demonstrate the link between cognitive load and unconscious behavior by requiring only a mouse click to enter the correct answer. Therefore, moving the mouse was not necessary to play this $n$-back game other than stopping the experiment (a stop button is located at the bottom). This means that moving mouse behavior is not relevant to the game completion. As shown in Fig.~\ref{img:gui_comparison} (b), the GUI included no buttons to indicate the position and audio matches. The game only lets participants click the left and right buttons of a mouse to indicate position or audio matches. For example, participants need to click on the right mouse button when an auditory cue does match with $n$ steps before. In the GUI, the strings `Position Match' and `Audio Match' give feedback from participants' answers, and those strings are not buttons to indicate answers. When responses took place with the left button on the mouse and it does not match with the previous $n$-step position cue, the color of `Position Match' characters turns to red giving negative feedback to the participant. The demo video of playing the dual $n$-back game used for this study can be viewed through the external video: https://youtu.be/ZsUDPl5Yr88.

\subsection{Mouse Data Preprocessing} \label{data_extraction}

\begin{figure}
\centering
\includegraphics[width=0.95\linewidth]{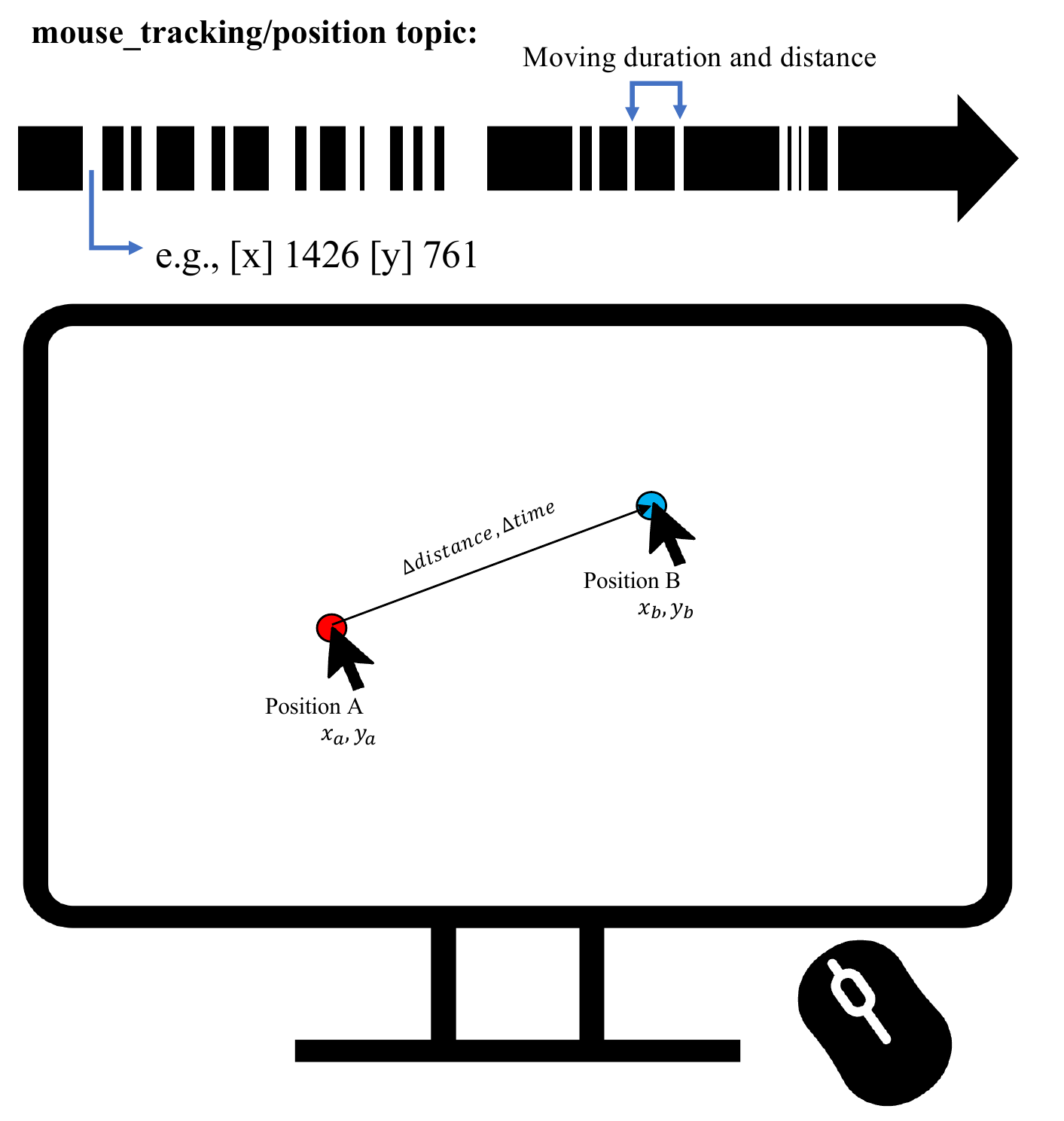}
\caption{(Top) Mouse data extraction method: The mouse\_tracking/position topic reproduces the experimental environment and indicates where the mouse position was during the same time period. Motion time and distance were extracted by using the difference between topics. (Bottom) Mouse data features from two positions.}
\label{img:mouse_features}
\Description{The figure at the top represents how the mouse-related ROS topic consists of time and coordinates in the screen. A segment at the front has an example of a saved coordinate in the topic. Two adjacent segments have a caption that moving duration and distance are retrieved from the segments. The bottom figure represents how two different mouse positions were used to extract the distance and time changes during the experiment.}
\end{figure}

The affective dataset provides the ROS-topic (\textit{/mouse\_tracking/position}) to retrieve mouse behaviors with time information when participants move their mouse during the user study. Experimental data was used when participants played the dual $n$-back game, which was about 60 seconds. 
As shown in the top of the Fig.~\ref{img:mouse_features}, when the ROS environment reproduced the experiment environment, the user's mouse position data and movement were extracted with ROS time information following saved discrete topics that have $x$ and $y$ positions with time data indicating the participants moved the mouse cursor to the saved position. 
We extracted three pieces of information by ROS time and mouse position information: movement frequency, movement duration, and pixel position change. If two positions were extracted from the ROSbag file, as shown in the bottom of the Fig.~\ref{img:mouse_features}, moving frequency is recorded. Movement duration was extracted from ROS time recorded with the position data. The pixel position changes were calculated by the distance between two mouse cursor location data. 

\section{Analysis} \label{chap:analysis}

The 27 participants were chosen among 30 sets excluding the damaged three sets of data.
Firstly, extracted mouse trajectories were drawn in images to show how the participants' unconscious movements are distributed. The trajectories show to what extent individuals move the cursor during the experiments. All mouse trajectories are presented in the supplementary video. One participant did not show any movements during the dual 3-back game, while three participants showed a fixed mouse in the 1-back game. 

\begin{figure}
\centering
\includegraphics[width=0.98\linewidth]{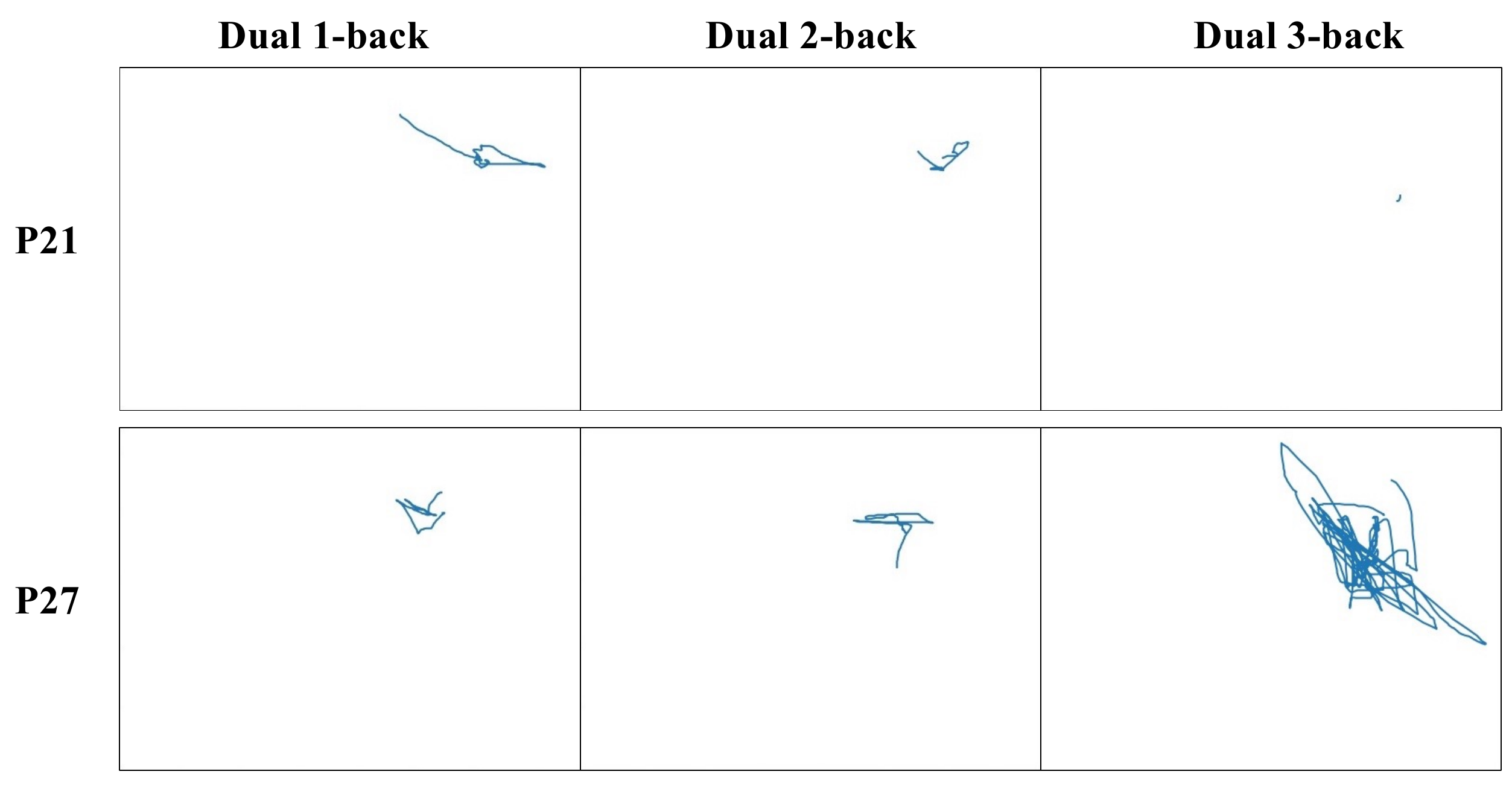}
\caption{Mouse trajectories of two participants were combined. The participant 21 showed decreased movements when the levels of dual $n$-back game increased, while the other participant showed more active movements as the levels increased.}
\label{img:mouse_comparison}
\Description{Mouse trajectories of two participants were combined. The participant 21 showed decreased movements when the levels of dual n-back game increased, while the other participant showed more active movements as the levels increased.}
\end{figure}

Fig.~\ref{img:mouse_comparison} shows mouse trajectories of two participants in the dataset. The participant P21 showed decreasing movements throughout the experiment (see the top three images). In contrast, the participant P27 showed increasing movements as the level increases (see the bottom three images). A spotted point in trajectory drawings is that human unconscious movements appear differently. The mouse-related behaviors may appear in the opposite direction as the level increases.

We used one-way repeated measure ANOVA (rmANOVA) \cite{benedetto2011driver, singh2013analysis} to prove variance among levels statistically. The IBM SPSS Statistics software tested 81 sets of mouse tracking. The one-way ANOVA test aims to distinguish three individual levels that are statistically different in 95\% confidence interval. First, we analyze data descriptively, means and standard deviation values. Each mouse measure is tested through the ANOVA to find which measure is significant. If any behavior pattern is confirmed, the three levels are compared with the valid method. 

Table \ref{tab:mean_std_result} shows mean and standard deviation values of measures. Mouse movement frequency rates were similar among levels (Low: 66.48, Medium: 66.48, High: 67.40 counts). 
The pixel changes in the high level were the highest. However, the standard deviation shows the variance among participants was also high. The descriptive result in pixel changes shows some participants did not move the mouse wide, while others changed the cursor broadly. Movement duration was longer in the high-level game.
The rmANOVA compared the three levels of mouse measure. Table \ref{tab:mouse_anova} represents that the movement duration and pixel changes did not pass the test. Measuring movement frequency showed the significance to differentiate three game levels (H1a accepted: $F(2,52)=3.673$, $p$ \textless .05). The ANOVA test also compared the frequency data between levels, represented in Table \ref{tab:mouse_anova_between}. The one-way ANOVA showed the low and high workload showed a significant difference, while frequent movements at medium level cannot be distinguished from the previous or increasing workload. The test result means that participants unconsciously reacted differently in low and high levels shown in mouse movements. Also, the result could signify that participants consecutively moved the mouse more frequently as cognitive level increases.

\begin{table}
\caption{A descriptive analysis result of mouse behavior.}
\begin{center}
\begin{tabular}{cccc}
    \toprule
    \textbf{Measures} & \textbf{Levels} & \textbf{Mean} & \textbf{Std.} \\
    \midrule
    \multirow{3}{*}{Movement Frequency (count)} 
    & Low & 66.48 & 11.164 \\
    & Medium & 66.48 & 10.785 \\ 
    & High   & 67.40 & 11.132 \\
    \midrule
    \multirow{3}{*}{Movement duration (milliseconds)} 
    & Low & 3.27 & 3.478 \\
 & Medium & 3.56  & 4.896  \\
 & High   & 6.03  & 9.554  \\
    \midrule
    \multirow{3}{*}{Movement position change (pixels)} 
    & Low & 14.70 & 28.496 \\
 & Medium & 13.43  & 20.676  \\
 & High   & 29.90 & 50.688 \\
    \bottomrule
\end{tabular}
\end{center}
\label{tab:mean_std_result}
\Description{The table represents the descriptive analysis results of mouse movement frequency, movement duration, and position change.}
\end{table}

\begin{table}
\caption{The rmANOVA test results of movement frequency, movement duration, and pixel position change.}
\begin{center}
\begin{tabular}{ccl}
        \toprule
        \textbf{Measure} & \textbf{F(2,52)} & \textbf{Significance} \\
        \midrule
        Movement Frequency (count) & 3.673 & \textbf{.032 ($\boldsymbol{p}$ \pmb{\textless} $\boldsymbol{.05}$)} \\
        Movement duration (milliseconds) & 1.699 & .193 ($p$ \textgreater $.05$) \\
        Movement position change (pixels) & 1.711 & .191 ($p$ \textgreater $.05$) \\
        \bottomrule
    \end{tabular}
\end{center}
\label{tab:mouse_anova}
\Description{The table represents the rmANOVA test results of mouse movement frequency, movement duration, and pixel position change. Movement frequency showed statistical significance.}
\end{table}

\begin{table}
\caption{Movement frequency changes: ANOVA test results between levels.}
\begin{center}
\begin{tabular}{ccl}
            \toprule
            \textbf{Level comparison} & \textbf{F(1, 26)} & \textbf{Significance} \\
            \midrule
            Low \& Medium  & .000  & 1.000 ($p$ \textgreater $.05$) \\
            Low \& High    & 5.286 & \textbf{.030 ($\boldsymbol{p}$ \pmb{\textless} $\boldsymbol{.05}$)}  \\ 
            Medium \& High & 4.016 & .056 ($p$ \textgreater $.05$)  \\
            \bottomrule
        \end{tabular}
\end{center}
 \label{tab:mouse_anova_between}
 \Description{Mouse movement frequency between levels is analyzed. The low and high levels showed statistical significance.}
\end{table}

\section{Data Validation}
We validate the hypothesis that position changes in mouse usages have a relationship with the levels of cognitive load, through the subjective responses and behavioral measures. The NASA-TLX self ratings and eye-related behavior were obtained together from the affective dataset \cite{jo2020rosbag}. The dataset provided a recorded video stream through ROS topic \textit{/camera/color/image\_raw} with mouse tracking data at the same time. The self-rating responses were given along with ROSBag files.

\subsection{NASA-TLX}

\begin{figure}
\centering
\includegraphics[width=0.9\linewidth]{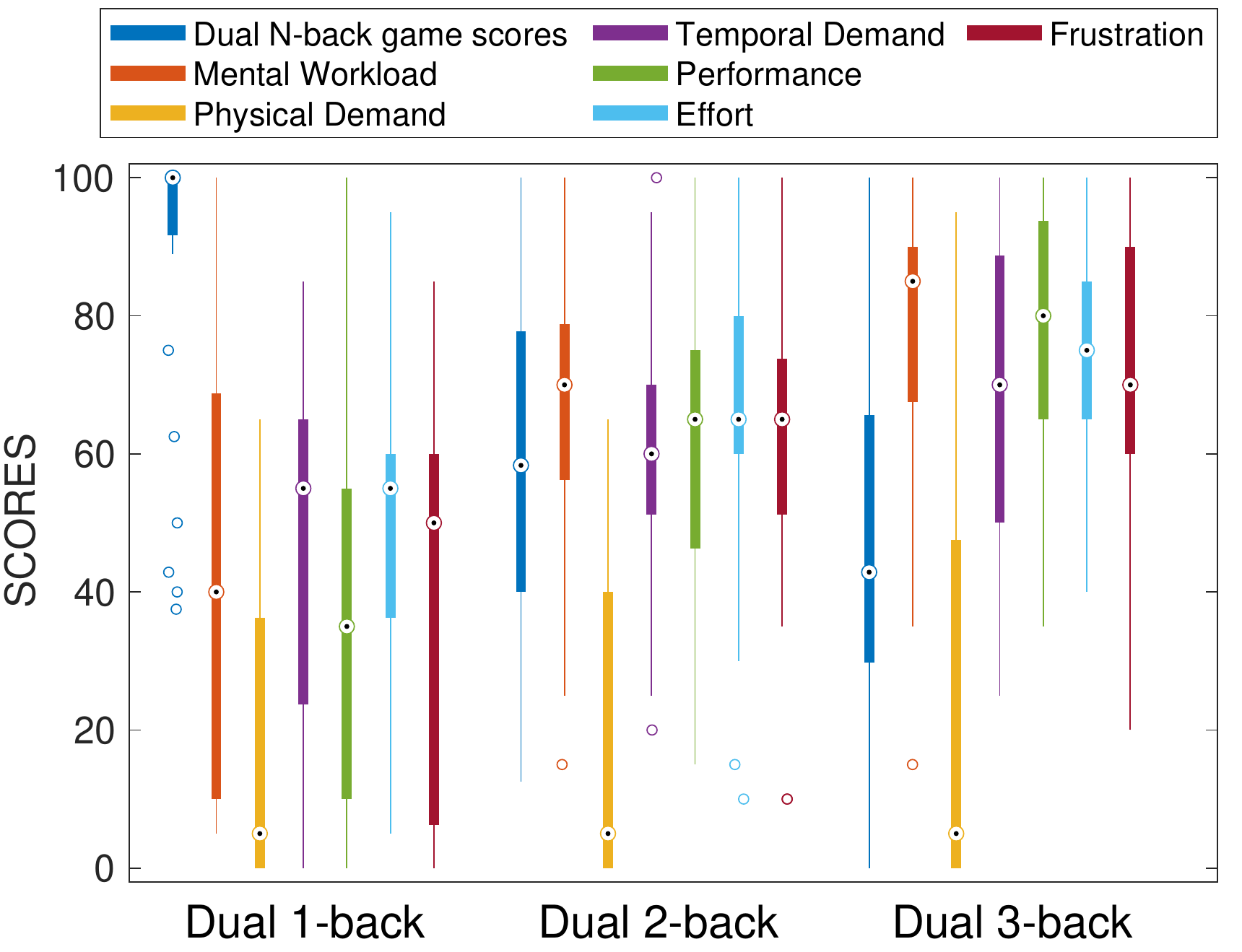}
\caption{Dual $n$-back test and NASA-TLX results: The white circles indicate the average, and the color bars show the data distribution.}
\label{img:nasa-tlx}
\Description{The dual n-back scores and NASA-TLX ratings are represented in three box plots. The ratings of the dual 1-back have mean values between 37 and 55 on mental workload, temporal demand, performance, effort, frustration. The ratings of the dual 2-back have between 60 and 70 on mental workload, temporal demand, performance, effort, frustration. The ratings of the dual 3-back have between 65 to 80 on mental workload, temporal demand, performance, effort, frustration. The dual n-back game scores decrease from the dual 1-back to the dual 3-back. The mean values of the physical demand ratings are similar, but the range of ratings is more expansive as the level increases.}
\end{figure}

We examined the dual $n$-back score and the NASA-TLX results to determine the difference between the three levels of cognition. Fig. \ref{img:nasa-tlx} shows the game score and NASA-TLX ratings of 27 participants. $n$-back game scores decreased as the game level rises, as the blue color bar shows in the figure. The average values of mental workload, temporal demand, performance, effort, and frustration increased as the game became more challenging. Table \ref{tab:nasa-tlx-ratings} presents comparison results of ratings gathered in NASA-TLX answers and game scores. The physical demand ratings were not significant between levels, which means individuals did not observe physical needs. The main difference between mouse data and the NASA-TLX is that the self-rating showed the difference between low and medium workload levels.

\begin{table}
\caption{The rmANOVA test results on NASA-TLX ratings: Most of ratings, other than physical demand, showed significance.}
    \label{tab:nasa-tlx-ratings}
\begin{center}
\begin{tabular}{ccl}
    \toprule
    \textbf{Rating} & \textbf{F(2,52)} & \textbf{Significance}  \\
    \midrule
    Game score      &  35.200& \textbf{.000} ($\boldsymbol{p}$ \pmb{\textless} $\boldsymbol{.05}$) \\ 
    Mental demand   & 35.490 & \textbf{.000} ($\boldsymbol{p}$ \pmb{\textless} $\boldsymbol{.05}$) \\
    Physical demand & 2.850  & .067 ($p$ \textgreater $.05$) \\ 
    Temporal demand & 14.137 & \textbf{.000} ($\boldsymbol{p}$ \pmb{\textless} $\boldsymbol{.05}$) \\ 
    Performance     & 28.939  & \textbf{.000} ($\boldsymbol{p}$ \pmb{\textless} $\boldsymbol{.05}$) \\
    Effort          & 18.838 & \textbf{.000} ($\boldsymbol{p}$ \pmb{\textless} $\boldsymbol{.05}$) \\ 
    Frustration     & 31.824 & \textbf{.000} ($\boldsymbol{p}$ \pmb{\textless} $\boldsymbol{.05}$) \\ 
    \bottomrule
\end{tabular}
\end{center}
\Description{NASA-TLX ratings of three different levels of the dual n-back games are analyzed with the rmANOVA test. Most of ratings, other than physical demand, showed statistical significance.}
\end{table}

\subsection{Eye blinking Duration}
\begin{figure}
\centering
\includegraphics[width=0.8\linewidth]{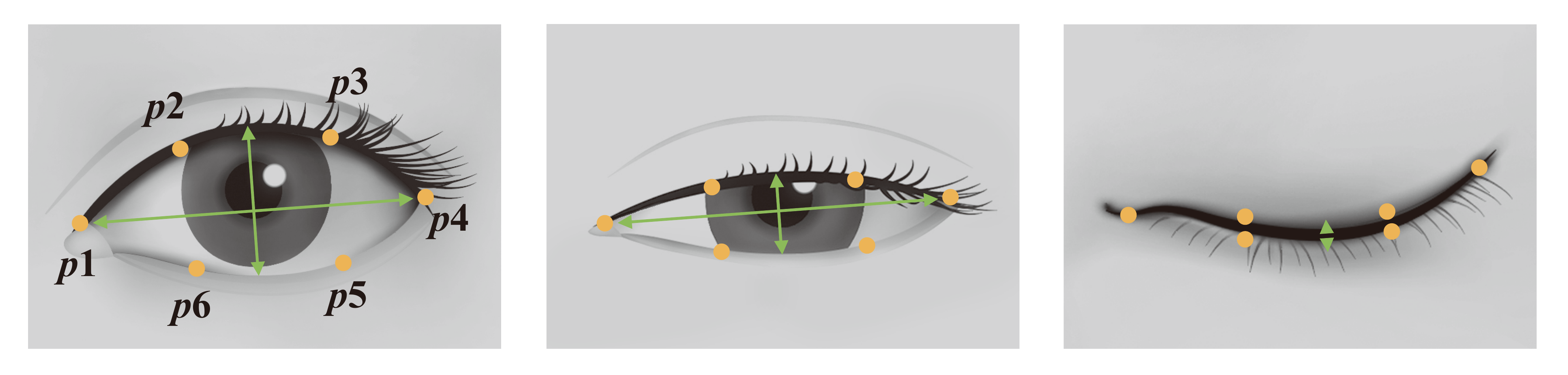}
\caption{Eye Aspect Ratio (EAR) calculation in Eq. \ref{ear_equ}. Feature points (from p1 to p6) from an eye can be retrieved from recorded videos.}
\label{img:feature_extraction_eye}
\Description{The figure extracting six feature points for the eyes is shown in the figure. The change was depicted until the eyes were closed.}
\end{figure}

Eye blinking is one of the unconscious behaviors as mentioned in Section \ref{literature}. We analyzed the eye blinking frequency and blinking duration, because they are effective indicators for measuring cognitive load \cite{benedetto2011driver}. The used dataset captured participants' eye-blinking behavior with a front-facing camera within the experimental environment. In order to detect the blinking of the eyes, we utilized an open-source library, OpenPose \cite{openposegithub}, that uses deep-learning estimation to capture seven landmarks on each eye region. Eye blinking was detected by 
\begin{equation} \label{ear_equ}
Eye\ Apsect\ Ratio~(EAR)= \frac{\left \| p2-p6 \right \| + \left \| p3-p5 \right \|}{2\left \| p1-p4 \right \|},
\end{equation}

\noindent using the landmark detected in the eye area \cite{cech2016real}. Each landmark used in the eye aspect ratio (EAR) calculation is depicted in Fig. \ref{img:feature_extraction_eye}. If the value of the EAR exceeds the given threshold, 0.2, the eye blinking events were recorded. The eye closing time of participants was analyzed in milliseconds. Among the participants, 16 sets of data were acceptable to be analyzed due to glares on eyeglasses as mentioned in Section \ref{dataset}.

\begin{table}
\caption{Eye blinking behavior data: An rmANOVA result of blink frequency and blink duration.}
    \label{table:total_anova}
\begin{center}
\begin{tabular}{lcc}
    \toprule
    \textbf{Measure} & \textbf{F(2, 30)} & \textbf{Significance} \\
    \midrule
    Blink frequency        & 0.038 & $.962$ ($p$ \textgreater $.05$) \\
    Blink duration         & 3.50  & \textbf{.043 ($\boldsymbol{p}$ \pmb{\textless} $\boldsymbol{.05}$)} \\
  \bottomrule
\end{tabular}
\Description{Eye blinking behavior data are also analyzed with the rmANOVA analysis. Eye blinking duration showed statistical significance.}
\end{center}
\end{table}

\begin{table}
    \caption{Eye blinking duration: ANOVA test results between levels.}
    \label{tab:ebd_anova_between}
\begin{center}
\begin{tabular}{ccc}
            \toprule
            \textbf{Level comparison} & \textbf{F(1, 15)} & \textbf{Significance} \\
            \midrule
            Low \& Medium  & 1.559 & .231 ($p$ \textgreater $.05$) \\ 
            Low \& High    & 6.699 & \textbf{.021} ($\boldsymbol{p}$ \pmb{\textless} $\boldsymbol{.05}$)  \\ 
            Medium \& High & 2.015 & .176 ($p$ \textgreater $.05$) \\ 
            \bottomrule
\end{tabular}
\Description{Eye blinking duration data between levels are also analyzed. The low and high levels showed statistical significance.}
\end{center}
\end{table}

One-way rmANOVA is utilized with eye blinking frequency and duration to examine if three separate levels are distinctive. Among two eye-related measures, eye-opening duration was also feasible to detect different levels of mental load ($F(2,30)=3.50$, $p$ \textless .05), as shown in Table \ref{table:total_anova}. As the same procedure as we analyzed the valid measure of mouse behavior, we compared the duration of the levels. Table \ref{tab:ebd_anova_between} shows that the low and high levels are different from the eye-opening cues of participants. The analysis results from eye blinking duration comply with mouse behavior analysis, especially from mouse moving frequency.

\section{Discussion}
We determined that unconscious mouse actions are correlated to human cognitive workload. The statistical analysis confirmed that the frequency of mouse movement representing how redundant behavior distinguishes the different workload levels. The moving duration and changing the location of mouse usage were incongruent. We validated the mouse-related data analysis with the self-questionnaire answers related to human workload, the NASA-TLX, and eye blinking behavior. Eye blinkings were analyzed with a smaller portion of the dataset due to the image distortion and glitches on the eyeglasses of participants. Interestingly, however, differentiating human cognitive workload by two kinds of unconscious behaviors was valid. Two measures, mouse movement frequency and eye blinking duration showed their significance in distinguishing low and high mental load levels. The self-ratings support the dual $n$-back game successfully derived the different cognitive load levels. 

Three questions to consider regarding the dataset existed. First, the dual $n$-back game allows users to click mouse buttons, but the number of clicking varies depending on the game level. Trials occur 20 times for 60 seconds in the dual 1-back and 18 possible events in dual 3-back. The mouse data analysis in Section \ref{chap:analysis} showed the almost equivalent frequency among the levels, which concluded that the number of trials does not correlate with mouse cursor changes caused. Second, considering that the mouse-related topic is recorded when the cursor moves, the ROS system can also save slight movements. Click events may generate slight movements when the participants responded, but the mouse trajectory data were recorded randomly than regularly or linearly. However, recorded mouse movements with little change were written when the test subject moved the mouse once or twice during the experiment. Therefore, the movement that may occur when the mouse is pressed has not been considered. Finally, the experimental GUI had a stop button that could also derive task-irrelevant mouse trajectories. However, trajectories among levels did not record any trace around the area of the stop button. The stop button did not influence on experiments.

This study assumed that unconscious mouse behavior is derived from cognitive states. However, affective states may influence the participants to make redundant behaviors with the mouse. The cognitive workload task in the affective dataset provides the NASA-TLX questionnaire but does not include the SAM Rating \cite{bradley1994measuring} considering the participants' affective states. Therefore, we may need to verify that the mental load causes the unconscious behavior.

\section{Conclusion and Future Work}
We concluded that unconscious mouse actions and human cognitive workload are correlated. This study hypothesized that unconscious mouse movements that are not related to given tasks differentiate different levels of human cognitive load. Analyzing mouse movement frequency confirmed that our hypothesis is valid; in other words, human cognitive load can be predicted by watching how many times individuals change the location of mouse unconsciously. However, the relationship between unconscious usage and the participants' affective state was not confirmed. As future work, the redundant mouse actions should be investigated whether affective states or cognitive states result in the actions.

\begin{acks}
This material is based upon work supported by the National Science Foundation under Grant No. IIS-1846221. Any opinions, findings, and conclusions or recommendations expressed in this material are those of the author(s) and do not necessarily reflect the views of the National Science Foundation.
\end{acks}

\bibliographystyle{ACM-Reference-Format}
\bibliography{main}

\end{document}